\documentclass[sort&compress,11pt,3p]{elsarticle}
\usepackage{lineno}
\usepackage{amssymb}
\usepackage{amsmath,epsfig}
\usepackage[latin1]{inputenc}
\usepackage{graphicx}
\usepackage[english]{babel}
\usepackage{xspace}
\usepackage{subfigure}
\usepackage{float}
\usepackage{multirow}
\usepackage{placeins}
\usepackage[percent]{overpic}%https://it.overleaf.com/project/5e977f5815bb070001f45cbchttps://it.overleaf.com/project/5e977f5815bb070001f45cbc
\usepackage{eurosym}
\usepackage{hyperref}
\usepackage[table,xcdraw]{xcolor}
\usepackage{array}
\usepackage{todonotes}
\newcounter{todocounter}

\newcolumntype{L}[1]{>{\raggedright\let\newline\\\arraybackslash\hspace{0pt}}m{#1}}
\newcolumntype{C}[1]{>{\centering\let\newline\\\arraybackslash\hspace{0pt}}m{#1}}
\newcolumntype{R}[1]{>{\raggedleft\let\newline\\\arraybackslash\hspace{0pt}}m{#1}}

\usepackage{color}

\usepackage{hyperref}

\makeatletter
\def\ps@pprintTitle{%
\let\@oddhead\@empty
\let\@evenhead\@empty
\def\@oddfoot{}%
\let\@evenfoot\@oddfoot}
\makeatother
\journal{Physics in Medicine and Biology}

\begin{document}
\begin{frontmatter}

\title{Performance of a low gain avalanche detector in a medical linac and characterisation of the beam profile}
\author[1]{T.~Isidori }
\author[2,3]{P.~McCavana }
\author[2,3]{B.~McClean }
\author[3]{R.~McNulty }
\author[1]{N.~Minafra}
\author[3]{N.~Raab }
\author[3,4]{L.~Rock }
\author[1]{C.~Royon}

\address[1]{
Department of Physics \& Astronomy,
University of Kansas,
Lawrence, KS 66045, USA
}
\address[2]{St. Luke's Hospital, Rathgar, Dublin 6, Ireland}
\address[3]{School of Physics, University College Dublin, Belfield, Dublin 4, Ireland}
\address[4]{Beacon Hospital, Sandyford, Dublin 18, Ireland}

\begin{abstract}

Low gain avalanche detectors can measure charged particle fluences with high speed and spatial precision, and are a promising technology for radiation monitoring and dosimetry. 
A detector has been tested in a medical linac where single particles were observed with a time resolution of 50\,ps.
The integrated response is similar to a standard ionising chamber but with a spatial precision twenty times finer, and a temporal precision over 100 million times better, with the capability to measure the charge deposited by a single linac pulse.
The unprecedented resolving power allows the structure of the $\sim 3\,\mu$s linac pulses to be viewed and the 350\,ps sub-pulses in the train to be observed.
\end{abstract}

\end{frontmatter}
%\linenumbers

\section{Introduction}

Dosimetry is essential in radiotherapy for understanding linac performance and to monitor the dose delivered to the patient.
Especially for small-field dosimetry, a high spatial precision is needed to distinguish between irradiation of the tumour and the surrounding healthy tissue and is particularly important in intensity-modulated radiation therapy or microbeam radiation therapy~\cite{grotzer}, where micron precision is desirable.
Time performance becomes important in dynamic environments when either the target is in  motion or the dose delivered varies with time.
With the advent of FLASH radiotherapy and the ability to delivery ultra-high radiation doses in fractions of a second~\cite{LEMPART201940}, there is a need for improved monitoring techniques~\cite{ashraf2020dosimetry}.
Detailed studies of the dosimetry and an understanding of the radiobiological mechanisms of Flash treatments would be greatly assisted by new detectors with
high temporal and spatial resolution.  

The standard monitoring tool in a clinical environment is the ionisation chamber, which
has a spatial precision of a few mm and a response time of about a second.
Combining several devices into a 2D array can allow the radiation profile to be mapped out, and studies have been performed using commercial devices~\cite{ALASHRAH2010181,anvari}, although these are limited in resolution and have significant dead-areas. 

The technology of choice for high spatial precision is silicon diodes, with dimensions from a few mm down to $50\,\mu$m, which can achieve resolutions of up to about $5\,\mu$m (for a recent review, see~\cite{Rosenfeld_2020}).
Individual diodes can be combined to create 2D arrays that have better resolution and granularity than ion-chamber arrays: an
example is the commercial SRS MapCHECK device consisting of 1013 diodes that cover a square of side 7.7\,cm~\cite{srsmap}.
Higher spatial precision and much greater granularity can be obtained by designing the diode array from scratch as strips or pixel implants in a single wafer of silicon.
Such silicon-strip or pixel detectors have routinely been used in the field of particle physics over the last 30 years, and are capable of detecting individual charged particles in  the dense high-radiation environment of the Large Hadron Collider (LHC)~\cite{ELBA_vertical,CMS:2667167,Collaboration:2623663}, with an almost 100\% efficiency.
These detectors have now become the baseline choice for many commercial activities \cite{TREMSIN2020106228,PROCZ2019106104} and have found several applications in dosimetry (e.g. \cite{Manolopoulos_2009,BOCCI201298,BISELLO201585,Alnaghy}), due to their high spatial and temporal resolutions.
Small pixel sizes give near-photographic resolution: the Medipix device~\cite{ROSENFELD2020106483}, an array of 256x256 pixels, each $55\,\mu$m square, can be used for X-ray imaging as well as charged particle detection.

One drawback of silicon devices for clinical settings has been their radiation hardness. 
However, significant advances have been made in the last fifteen years, driven by the necessity of operating in the intense radiation environment of the LHC.
Today, such detectors can be manufactured capable of withstanding up to $5\times10^{15}$ 1\,MeV neutrons per cm$^2$ (about 200\,kGy)~\cite{8331152,FERRERO201916}.

Time resolved measurements in radiotherapy have generally been limited to collection times in the ms range. 
A typical measurement time would be 0.25\,s for scanned beam profiles and depth doses in water tanks. 
The requirement for finer resolved measurements arrived with the advent of dynamic dose delivery where the gantry and MLC leaves move continuously during a treatment. 
The 2D arrays developed to analyse these doses include the Delta$^4$ diode phantom~\cite{delta4}, which with the use of a trigger pulse, can collect individual pulse doses over tens of $\mu$s. 
Current interest in FLASH therapy has led to further refinement with investigations into the individual $\mu$s pulses.  
Recently, scintillator fibres were shown to achieve sub-$\mu$s resolution for X-rays from a clinical Linac~\cite{FAVAUDON2019162537} using Cerenkov light from a fused silica cylinder to view the $\mu$s pulse in a FLASH beam, though this suffered from a ringing artifact. Diamond detectors are also capable of ns resolution of X-ray pulses~\cite{LIU2017248}.

The signal collection time for typical silicon strip or pixel detectors is about 10\,ns, which when readout, for example by the Timepix4 chip~\cite{BALLABRIGA2020106271}, allows the leading edge to be determined with a resolution of 200\,ps.   
The fastest time resolution however for silicon sensors is obtained with low gain avalanche detectors (LGAD), which married with specialised electronics, can record charged particles with a typical precision of 30\,ps.

This paper describes the performance of a LGAD, of dimensions 2.9 x 0.5\,mm$^2$, exposed to an electron beam of an Elekta linac.
This is a radiation-hard sensor designed to measure the timing difference (30\,ps) between protons at the LHC, which are 1\,cm apart and travelling essentially at the speed of light. 
The small dimensions of the device and the fast response mean that it can measure fluence rates of up to 100 million particles per square millimeter per second, and is thus capable of detecting single electrons in the linac pulse.
This allows the time-structure of the pulse, usually approximated as a square wave, to be investigated in detail.  
Understanding the temporal deposition of dose is of interest to radiobiologists as well as to machine engineers in understanding the operation of
a linac.

The apparatus used in these studies is
described below.
Section \ref{sec:ELEKTA} gives an overview of the linear accelerator, while the LGAD is described in Section \ref{sec:Detector}
and the methodology is outlined in Section \ref{sec:testsStLuke}.
The data analysis and results are presented in Section \ref{sec:electron}, and conclusions are given in Section \ref{sec:conclude}.

\section {Apparatus and methodology}

\subsection{The Elekta linac}
\label{sec:ELEKTA}

\begin{figure}[!ht]
\centerline{\includegraphics[scale=0.185]{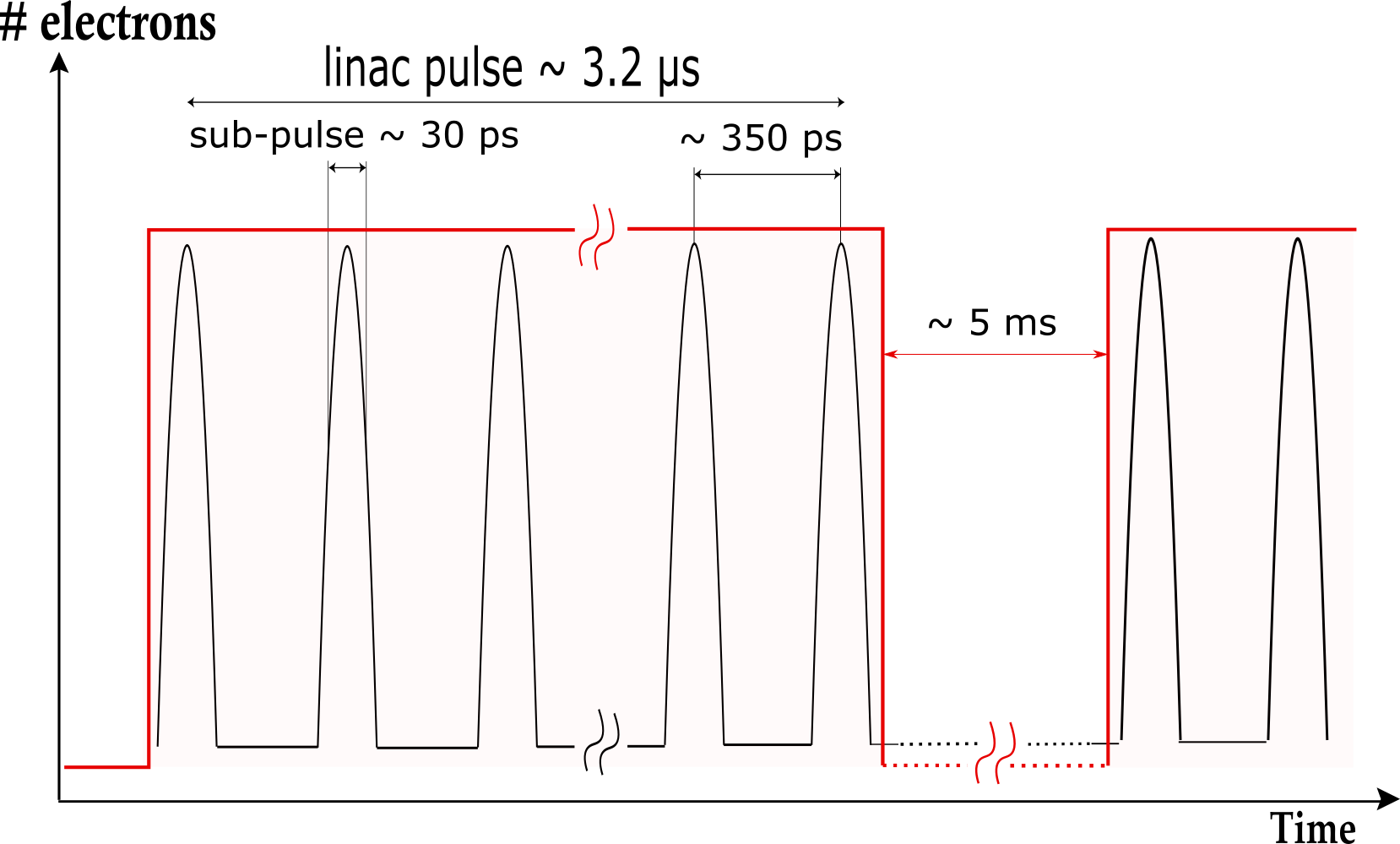}}
\caption{
Idealised profile of linac pulse \label{fig:sketch}
}
\end{figure}

An Elekta Precise (Elekta AB, Sweden) dual modality linear accelerator at Saint Luke's Hospital, Dublin, was used in these studies.
This has the capability to produce photon beams of energy 6\,MV with dose rates up to 600 MU/min and electron beams with energies between 4 and 18\,MeV and dose rates of 600 MU/min. 
The linac was decommissioned for clinical use several years ago but remains on a comprehensive maintenance and quality assurance program, including dose calibration, which allows its use for research purposes.
The research linac is also equipped with a multileaf collimator (MLC), an electronic portal imaging device (EPID) and a range of  applicators for electron beam collimation.
The accelerator operates at a pulse repetition frequency of 400\,Hz for photon beams and 200\,Hz for electron beam with each pulse length being about 3.2$\,\mu$s long.
Each of these pulse sequences contains thousands of 30\,ps sub-pulses separated by 350\,ps. 
The frequency of this fine structure is 2.856\,GHz.
An idealised schematic of the pulse is given in Fig.~\ref{fig:sketch}.

The dose was measured with a standard radiotherapy ion chamber. 
A PTW Semiflex Ionisation Chamber 31010
was operated at +400\,V.  
It consists of two layers of material (0.55\,mm PMMA and 0.15\,mm graphite) that encapsulate the sensitive volume of the detector.
The total active bulk is a cylinder of radius 2.75\,mm and height 6.5\,mm.
A dual channel PTW Tandem dual channel electrometer was used to read out the ion chamber with 10\,fA resolution. 
Its minimum measuring intervals of 10\,ms represent the bottleneck for the time required for every acquisition. 
The dead time between consecutive measurement is determined by the time (typically several seconds) needed for resetting the device.

\subsection{The low gain avalanche detector}
\label{sec:Detector}

LGADs are considered one of the most promising technologies currently available for fast and precise measurements of charged particles in high-rate environments.
The energy deposited by a particle passing through the sensor contributes to the creation of free charges (electron and holes) inside the active volume that drift towards the electrodes once an external electric field is applied. 
The motion of the charges generates a current that is amplified by a thin multiplication layer (hence the name LGAD), which is further amplified and shaped by the read-out electronics. 

The typical gain of a LGAD is in the range 5-20, significantly lower than that of an Avalanche Photo Diode (APD), whose gain can be up to 1000. 
The lower number of free charges generated per unit of deposited energy causes a reduction in the collection time, which in turns reduces the dead time after the avalanche process. 
In addition, the sensor has a lower dark current.
Thanks to these properties, it is possible to design a thin ($\sim$ tens of microns) sensor that produces a fast, low-noise signal.

\begin{figure}[!ht]
\centerline{\includegraphics[width=0.48\textwidth]{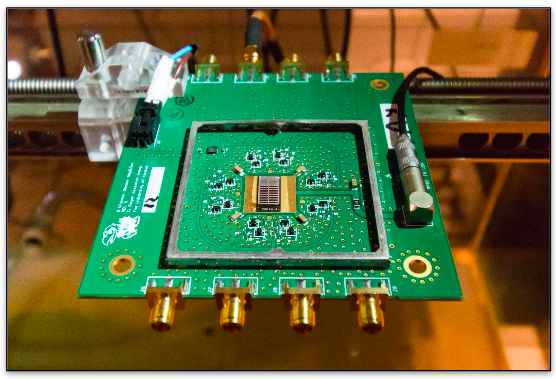}
\includegraphics[width=0.48\textwidth]{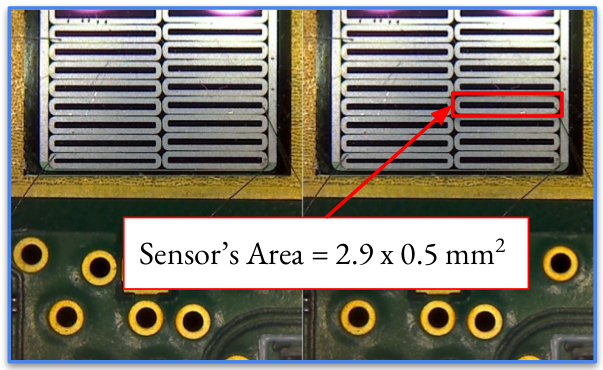}}
\caption{
Left: The KU-board mounted on the horizontal rail of an empty PTW 3D scanning water tank at St. Luke's Hospital, Dublin. 
Right: A close-up of the LGAD sensor aligned and glued to the KU-board. Only a single pixel (outlined in red) was used for these tests.}
\label{Fig:ufsd}
\end{figure}

The device used in the studies is displayed in Fig.~\ref{Fig:ufsd} and consists of a LGAD sensor mounted on a bespoke read-out board, developed at the University of Kansas (KU).
Only one of the pixels on the sensor was used with an active area of 2.9$\times$0.5\,mm$^{2}$.
The sensor was originally designed to operate inside experiments at the Large Hadron Collider~\cite{Albrow:2014lrm}.

The KU read-out board \cite{Minafra:2017nqc} is designed to host the sensor and the full amplification chain and is shown in
Fig.~\ref{Fig:ufsd}.
This compact and robust Printed Circuit Board (PCB) is designed to be easily configurable for use in different applications.
It consists of 8 identical two-stage trans-impedance amplifiers and a 20x20\,mm$^2$ pad that provides bias to the sensor up to $\sim$500\,V. 
Sensor pads can be wire bonded to one of the amplifiers.
Due to the high input impedance, the input capacitance has a large impact on the output signal; therefore, the sensor is placed at only few millimeters from the amplifier (to reduce the capacitance of the wire) and very high speed SiGe transistors are used (to reduce parasitic capacitance).
Previous tests with a similar configuration \cite{Berretti:2017qfr}, using minimum ionizing particles, showed a typical rise time of the order of $\sim$ 600\,ps, while the amplitude of the signal depends on the sensor properties, i.e. capacitance and gain.

The amplifier on the KU-board reads the current generated by the sensor, integrated on the capacitance of the sensor itself. 
The input impedance of the amplifier directly affects the integration time. In fact, a higher input impedance has a better signal-to-noise ratio (SNR) but a slower signal. 
However, a high impedance also requires more time to restore the baseline by removing the charge from the sensor. 
This presents a potential problem as subsequent signals will add to the charge already present and the amplifier will soon saturate. 
The choice of the input impedance leads to a compromise between higher rates and SNR.
In a similar way, since LGADs have a gain layer, higher gains lead to more charge collected and a higher SNR but a lower maximum rate.

The operating voltage for the detector was characterised during a previous test using a 180 GeV pion beam inside the LHC North Area Facility at CERN~\cite{Minafra:2017nqc}. 
The gain of the device depends strongly on the applied electric field.
A bias voltage too low will lead to inefficiencies in pulse shaping while, with a bias voltage too high, the sensor will generate too much charge during the passage of the particle, limiting the maximum rate of operation.
At high voltages (see Fig.13 in \cite{Minafra:2017nqc}), the detector achieves a time precision better than 25\,ps.  
However for the present studies, the detector was operated at a bias voltage of 150\,V, in order to give both good SNR and a fast response of the detector. 
At this working point, the expected time precision is about 50\,ps.

\subsection{LGAD measurements performed in the linac beam} 
\label{sec:testsStLuke}

\begin{figure}[!ht]
\centerline{\includegraphics[scale=0.3]{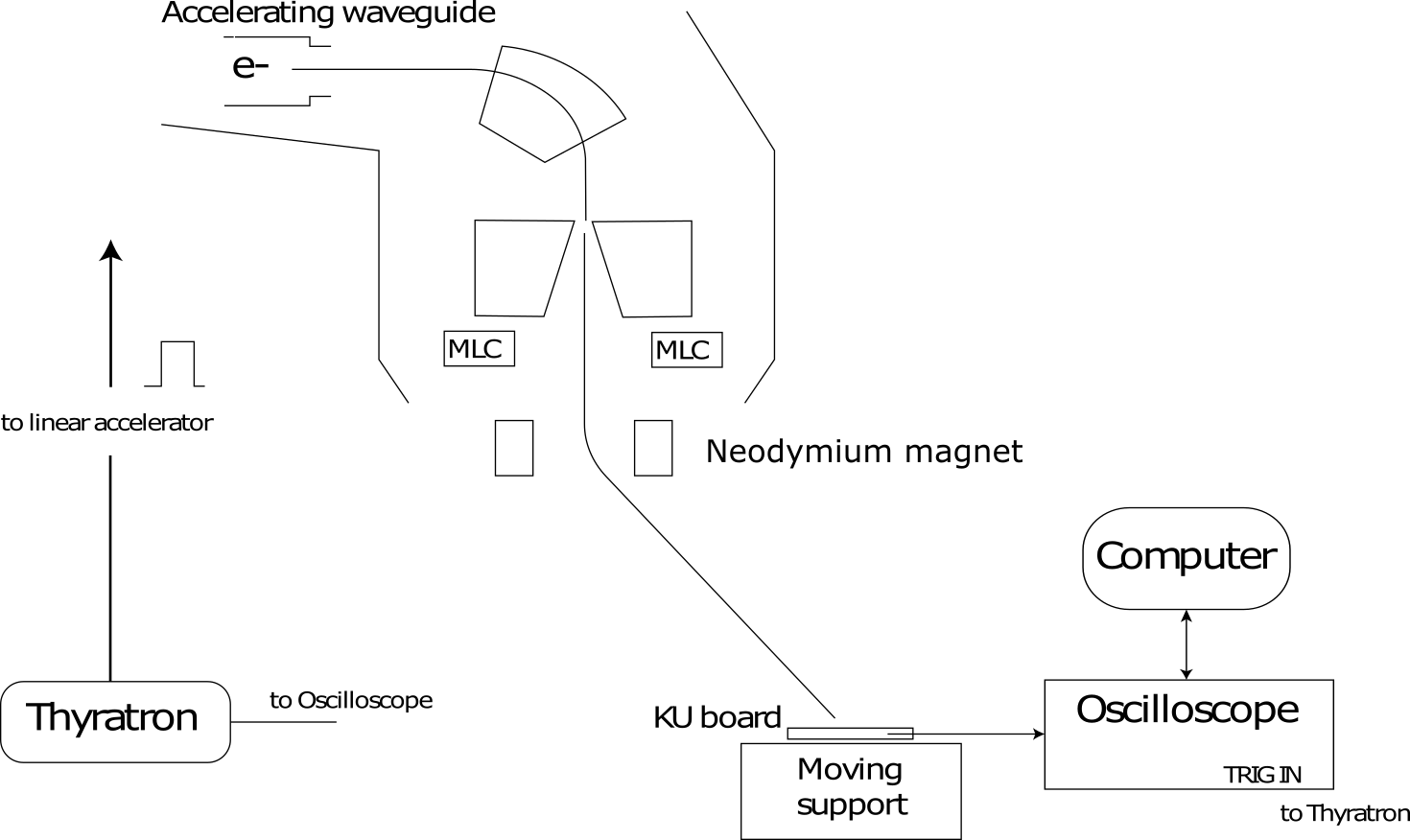}}
\caption{
The sketch shows the configuration used during the studies of the electron beam.
At the top is the linac head.  
The electron beam was deflected by the Neodymium permanent magnet that was placed at the exit of the multi-leaf collimator (MLC).
The KU-board was mounted on a moving support to scan the profile of the beam and connected to the oscilloscope.
The thyratron provided the trigger to the digitizer in the oscilloscope and the data were recorded on a computer.
}
\label{Fig:gantry}
\end{figure}

A 6\,MeV electron beam was used without an electron applicator and with the primary collimators set to 3x3\,cm$^2$.
A Neodymium N$^{40}$ permanent magnet was placed on the shadow tray, 12\,cm below the collimator faceplate, to bend the beam so that it spread along the horizontal axis.
This isolates the charged and neutral components of the beam, separates electrons with different momenta, and reduces the contamination of bremsstrahlung photons. 
The LGAD was mounted on the horizontal rail of an empty PTW 3D scanning water tank so its position could be changed remotely. 
The sensor was aligned vertically using the in-room positioning lasers and horizontally using the linac's light-field crosshair.
The moving support was used to scan the LGAD along the diverged beam and its output was recorded as a function of position relative to the source of radiation. 
Fig \ref{Fig:gantry} provides a simplified sketch of the setup used during the tests. 

The electrical current induced by the particle crossing the sensor was read out by the two stage amplification circuit producing a signal with a rise time of $\sim 600$\,ps and a total width of $5-10$\,ns. 
The signals were digitized using an Agilent\texttrademark DSO8104A Infiniium oscilloscope with a bandwidth of 1\,GHz and a sampling rate of 4\,GSa/s.
 
\begin{figure}[!ht]
\centerline{\includegraphics[width=.8\textwidth]{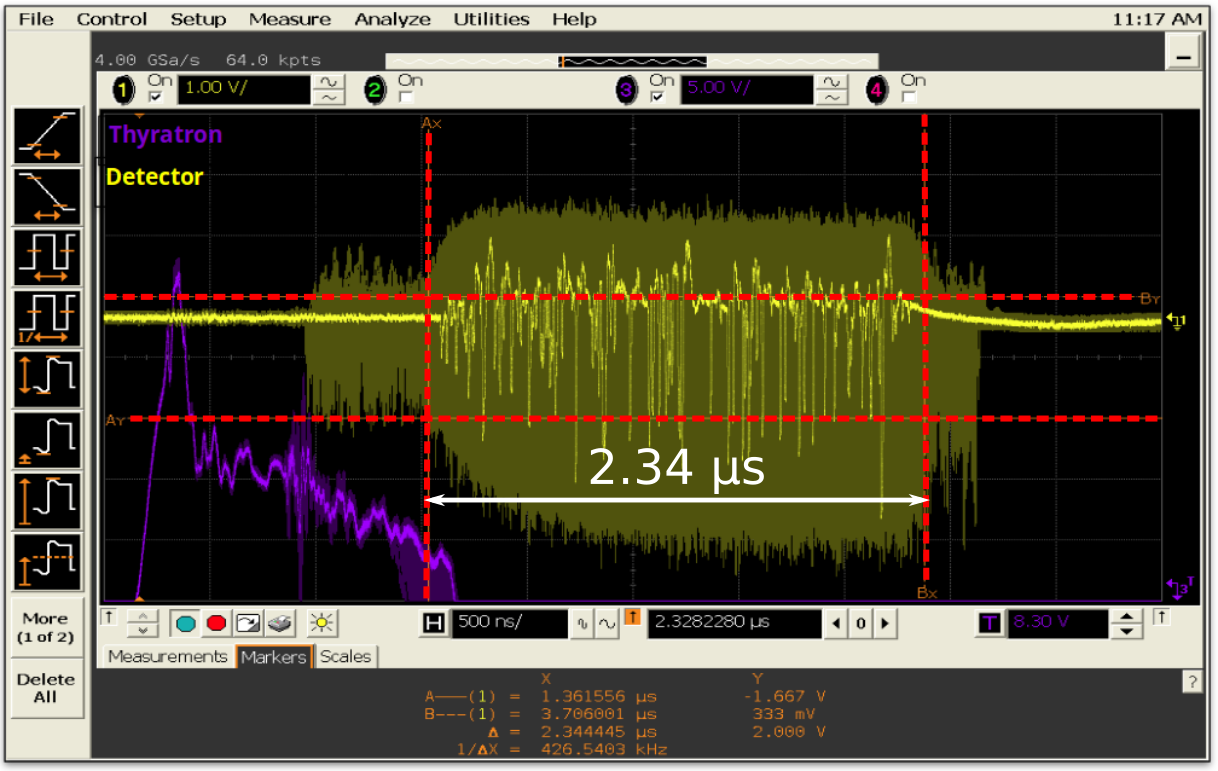}}
\caption{
Typical response of the LGAD (yellow) to one linac pulse as recorded by the oscilloscope.
The horizontal axis records time: The distance between the two red vertical scale-lines is 2.34$\,\mu$s.
The vertical axis records voltage: the distance between the two red horizontal scale-lines  is 2\,V.
The purple trace is the signal from the thyratron that provided the trigger to start the acquisition.
}
\label{Fig:oscilloscope}
\end{figure}

For each position of the sensor, the response of the detector to 200 pulses of the linac was recorded.
Fig.\ref{Fig:oscilloscope} shows a typical signal acquired by the oscilloscope (in yellow) during one pulse. 
The acquisition was triggered by a signal from the thyratron in the linac (in purple) corresponding to the electron injection inside the first acceleration stage.  
The data were subsequently analysed offline.

\section{Experimental results}
\label{sec:electron}
The data taken were analysed in three steps.  
First, the total charge collected by the LGAD per linac pulse was compared to that measured by an ion chamber.
Second, an algorithm was developed to identify individual charged particles traversing the detector, with a time precision of about 50\,ps.
Third, this precise time discrimination was used to investigate the structure of the linac pulse.

\subsection{Integrated charge collected by LGAD}

Fig.~\ref{Fig:oscilloscope} shows the data obtained during one linac pulse.
The beam is present from roughly the left-most red marker.  
The data were corrected for the DC offset,
which was measured by averaging data from 0.5 to 1.5$\,\mu$s before the pulse, while the RMS of the data in this region defined the intrinsic noise, $\sigma_{\rm noise}$.
The total charge collected during each linac pulse was found by integrating the signal.

\begin{figure}[!ht]
\centerline{\includegraphics[width=1\textwidth]{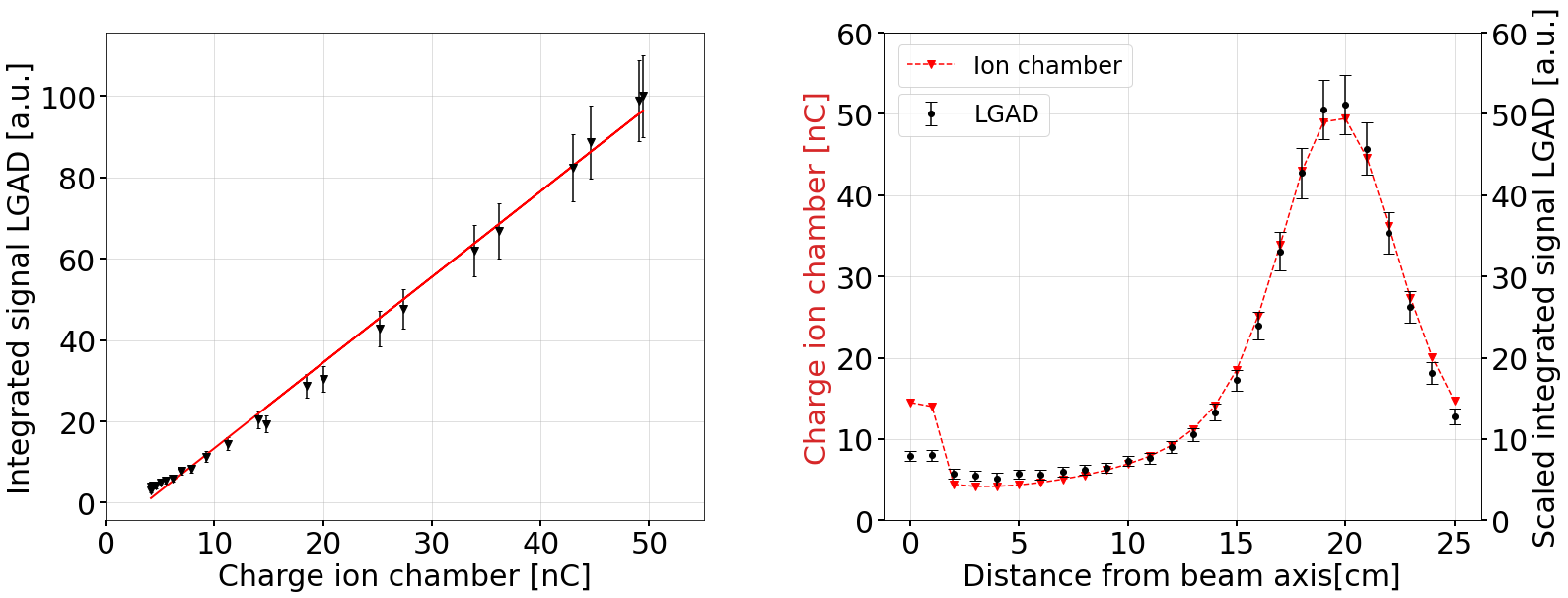}
}
\caption{
Left: Correlation between the charge in the ion chamber and the
average integrated signal in the LGAD measured for each linac pulse.
The red line shows the result of a straight-line fit.
Right: 
Charge measured with the ion chamber and average integrated signal recorded in the LGAD per linac pulse as a function of distance from the beam axis.
The error bars indicate the RMS of the signals obtained over 200 linac pulses. 
}
\label{fig:intcharge}
\end{figure}

The detector was moved laterally away from the central beam axis in 1\,cm steps and at each position the LGAD response was recorded for 200 linac pulses as was the response of an ion chamber (described in Sec.~\ref{sec:ELEKTA}) placed at the same location.
The average and RMS of the integrated signal in the LGAD is plotted in Fig.~\ref{fig:intcharge}(left) and compared to the ion-chamber response at all locations except those on the beam-axis.
A linear response is observed over most of the range, which is fit with a straight line.
The agreement between the two detectors shows that the LGAD gives equivalent results to the ion chamber, but with a time resolution of 3$\,\mu$s, rather than seconds.

Using the fit results, the LGAD response is scaled to that of the ion chamber.
Fig.~\ref{fig:intcharge}(right) 
shows the average and RMS of the integrated charge in the LGAD compared to the ion chamber
as a function of distance from the axis.
As expected, the responses are broadly similar for charged particles (the region $\ge2$\,cm from the beam) since both detectors have a highly efficient response to the passage of charged particles.
Differences here can be attributed to small
non-linearities in the read-out response~\cite{Minafra:2017nqc} as a function of deposited charge.

The principal difference between the LGAD and the ion chamber occurs under the (undeflected) beam position where there is a large flux of low-energy photons~\cite{Deng_2001}.
Differences in the responses of silicon detectors and ion chambers to photons have previously been observed: sometimes a larger signal is observed by the silicon and sometimes a smaller one~\cite{McKerracher_1999}.  
Since neither the LGAD nor the ion chamber directly detects photons, the response depends on how photons convert, both in the detectors and their housings.  
A complete understanding of the LGAD response to photons requires a full simulation of both the detector and the board on which it is mounted.  
In this paper, we limit further discussion to its response to charged particles.

The variation in the LGAD response, as reflected by the RMS, is due to the number of charged particles crossing the detector during a single linac pulse and the response of the electronics to variations in their temporal distribution.

\subsection{Single particle counting}
\begin{figure}[!ht]
\centerline{\includegraphics[width=1\textwidth]{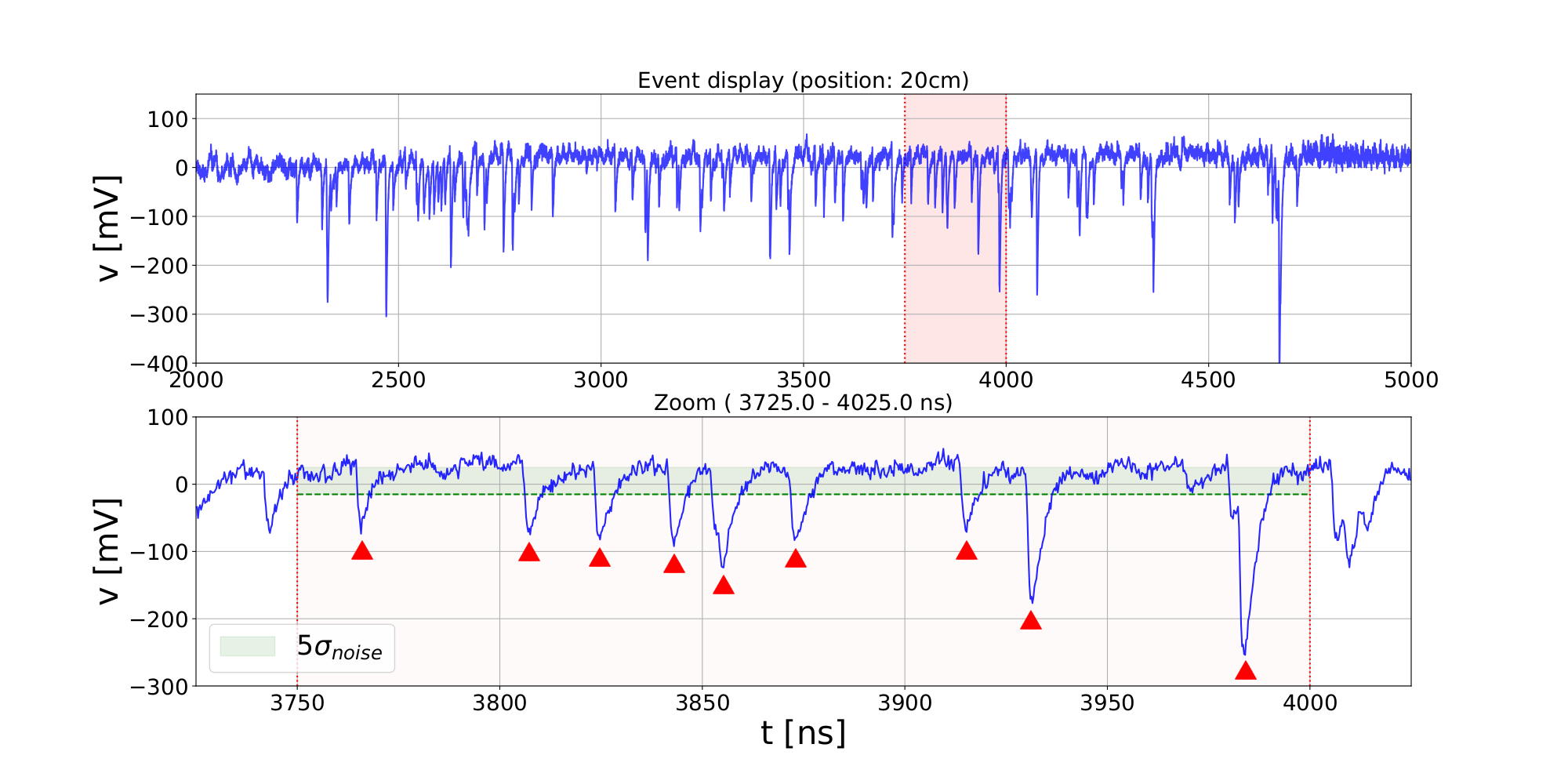}}
\caption{Top: Part of a typical LGAD signal obtained during one linac pulse; the abscissa shows the time elapsed since the trigger.
Bottom: Zoom on the selected area (highlighted in red). 
The width of the green band corresponds to five times the noise.
Isolated excursions from the baseline are indicated by red triangles and correspond to candidate charged particles passing through the silicon sensor, as identified by the algorithm.
}
\label{Fig:signal_detail}
\end{figure}

The ion chamber response time is typically a few seconds and requires several hundred linac pulses to obtain a measurement.
In contrast, the measurement for the LGAD presented above are for a single pulse.
However, even greater time precision is possible by identifying each charged particle that traverses the detector.
Fig.~\ref{Fig:signal_detail} shows details of the output signal for a single triggered event.  
In the zoomed display the signals resulting from charged particles crossing the detector are clear, and indicated in the figure by
red markers.

An automated algorithm was implemented to identify particles, and was designed such that it could be applied in real-time to give an instantaneous response in a clinical setting.
It proceeds in the following steps.
\begin{enumerate}
\item
The data are sent through a low-pass filter in order to reduce fluctuations due to noise.
\item
For each sample, a baseline is defined 
using the average of the channels in the
time interval 2-4.5\,ns before it.
This interval was chosen by inspection of the signal
pulses (Fig.~\ref{Fig:signal_detail}).  
A typical signal takes 1.2\,ns (5 samples) to go from the baseline to the maximum, after which the signal slowly decays.  To account for statistical fluctuations in
identifying the maximum channel, the window to
define the baseline starts 8 samples before the
putative maximum.
To account for statistical fluctuations
in the baseline level, the calculation is performed
over ten samples.
\item
A search is made for candidate channels greater than 
$5\sigma_{\rm noise}$ above the baseline
\item
The candidate channel is required to be the highest channel inside a window of $\pm 3$\,ns.
This time interval (corresponding to $\pm12$ samples) is
chosen by inspection of isolated signals, and is the typical time after which the signal is 50\% of the value at the peak.  
\end{enumerate}

The time-stamp for an individual particle crossing the detector, $T$, was taken as the time at which the leading edge of the signal crossed a threshold that was set to be 60\% of the distance from the baseline to the peak.
As discussed in Sec.~\ref{sec:Detector}, with the operating conditions chosen, the LGAD has a time precision of about 50\,ps.

Due to the high signal-to-noise of the LGAD, this algorithm is efficient for particles isolated in time.
However, when the time spacing between consecutive particles crossing the detector becomes significantly lower than the width of the signal pulse (around 10\,ns), the identification efficiency decreases.
For overlapping signals (e.g.\ between 4010-4020\,ns in Fig.~\ref{Fig:signal_detail}),
the reconstruction algorithm can fail to identify all particles.

\begin{figure}[!ht]
\centerline{
\includegraphics[width=1\textwidth]{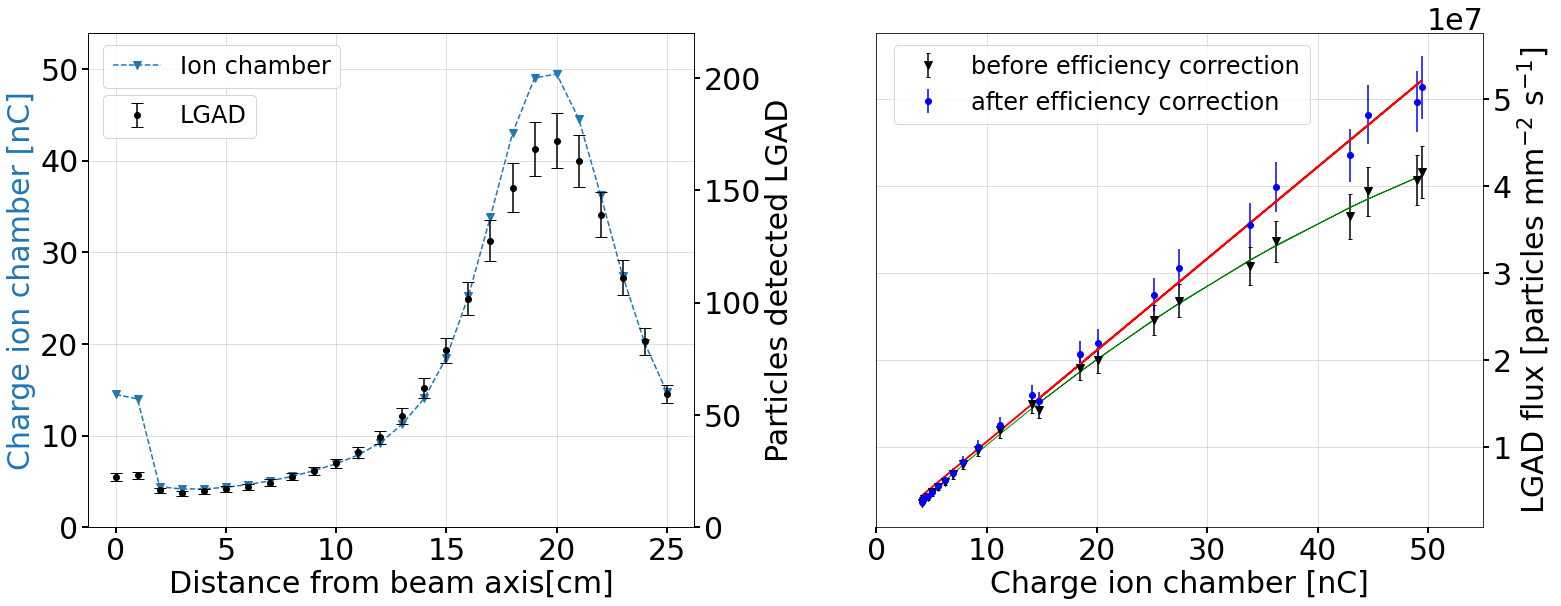}
}
\caption{Left: Comparison between the charge measured by the ion chamber and the average number of particles per pulse measured by the LGAD, as a function of detector position.
Right: Correlation between the ion chamber and LGAD, before and after an efficiency correction for high fluence rates.
The error bars indicate the RMS on the number of particles detected over 200 linac pulses.
}
\label{Fig:charge_comparison}
\end{figure}

The average number of particles found in a single linac cycle as a function of detector position is shown in the left panel of Fig.~\ref{Fig:charge_comparison} and compared to the ion chamber results.
It can be seen that the shapes are similar, although at high fluence rates, there is an inefficiency due to the intrinsic limit to the algorithm described above.
The inefficiency occurs when two particles pass through the detector in quick succession such that the signals overlap.  
The identification algorithm will always reject at least one when they are closer than 3\,ns, while when they are separated by more than 10\,ns, the algorithm is fully efficient.
On average, the algorithm fails when two particles pass through the detector within 6.5\,ns of each other.
The probability of this occurring is $\exp(-6.5\mu)$, where $\mu$ is the rate of particles going through the detector in GHz, and this defines an efficiency correction.
Fig.~\ref{Fig:charge_comparison} (right panel)
shows the correspondence between the charge recorded by the ion chamber and the number of particles observed in the LGAD, with and without this efficiency correction.
Before the correction, the response can be modelled with a second-order polynomial, while after, a linear response is observed.

The method of counting particles gives equivalent results to integrating the charge but now the detector is operating as a single-quantum detector: so long as the rate of charged particles passing through the detector is less than 100\,MHz, it can resolve single electrons with a time resolution of about 50\,ps, corresponding to the precision with which the leading edge can be measured.

\subsection{Characterisation of the linac beam}

\begin{figure}[!ht]
\includegraphics[width=1\textwidth]{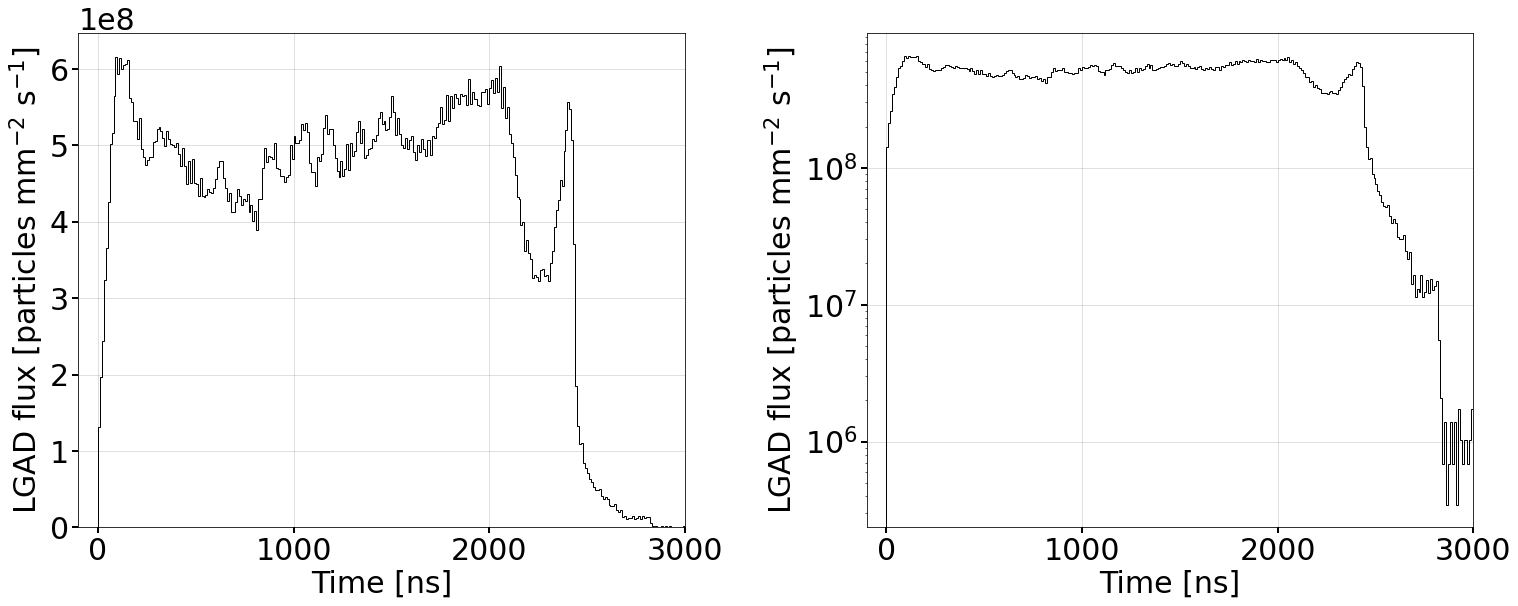}
\caption{
Particle flux measured by
the LGAD, summed over all positions, as a function of time on a linear (left) and log (right) scale.
}

\label{fig:beamshape}
\end{figure}

The excellent time resolution and linearity of the LGAD allows the dose delivered by the linac to be studied as a function of time within the pulse.
The number of particles recorded at all detector positions was summed and plotted as a function of time.
This allows the temporal profile of the beam intensity to be seen, and since the relative dose delivered scales with the charged particle flux, 
this identifies the fluctuations in the delivered dose during the linac pulse.
The result is shown in Fig.~\ref{fig:beamshape}.
It approximates to the 
idealised square pulse shown in Fig.~\ref{fig:sketch} but the precision of the LGAD allows a clear sub-structure to be observed.  
Furthermore, the width of the pulse is seen to be smaller than nominal at $2.85\pm0.01\,\mu$s.

Another feature that can be investigated is the train of sub-pulses that make up the microsecond pulse (see Fig.~\ref{fig:sketch}).
The linac operates with a radio-frequency of 2.856\,GHz so the sub-pulses (each expected to have a width of 30\,ps) are separated by 350\,ps.
 Given a resolution of about 50\,ps on the time-stamp for each particle crossing the detector, the sub-pulses should be visible so long as all other timing parameters within the linac are well aligned and have small jitter.  
Unfortunately, a large timing uncertainty was found between the thyraton trigger signal and the arrival of the linac beam (shown as the purple and yellow traces in Fig.~\ref{Fig:oscilloscope}).

\begin{figure}[!ht]
\includegraphics[width=1\textwidth]{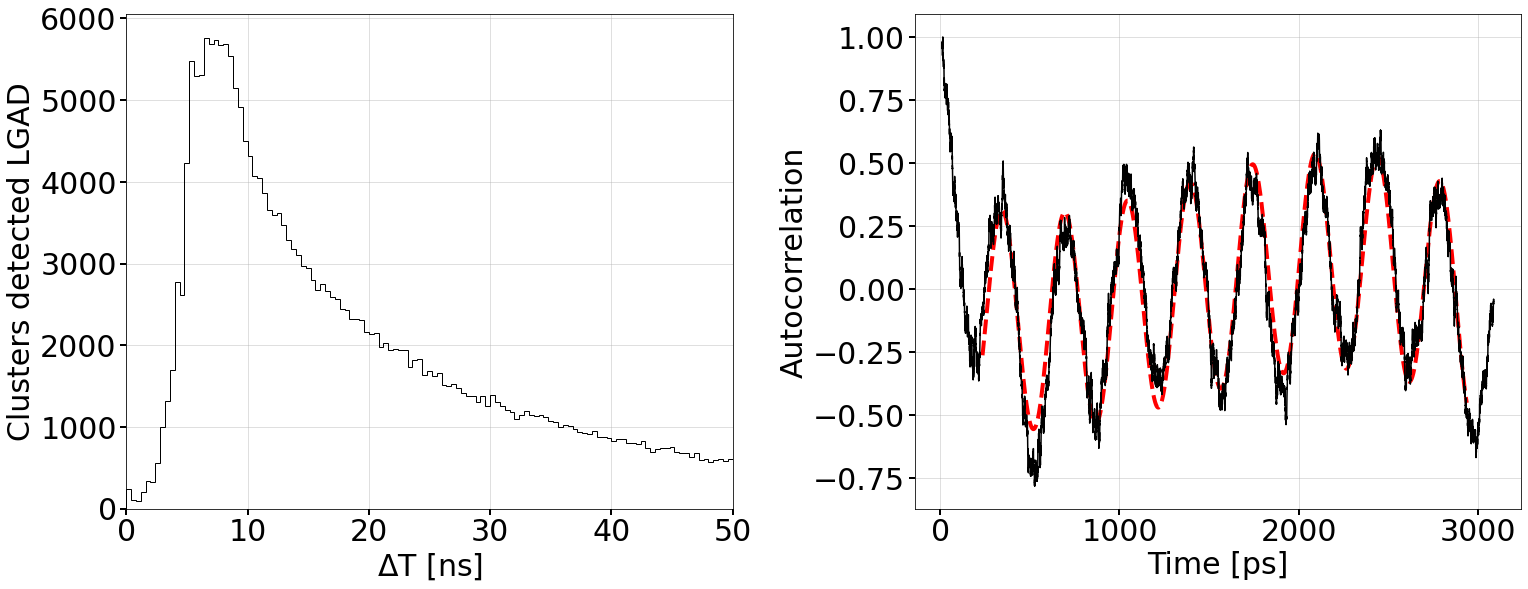}
\caption{Distribution of the time between consecutive particles in the beam.  
The left plot shows the full distribution. 
The right plot shows the autocorrelation fitted with a modulated sine function.  
}
\label{fig:pico}
\end{figure}

In the absence of a precise trigger and to reduce potential decoherence over several hundreds of ns, the sub-pulse structure was searched for using the distribution of the difference between the time-stamps for consecutive particles crossing the detector, $\Delta T$, which is shown in Fig.~\ref{fig:pico} (left).
For Poisson-distributed events, this distribution should be an exponential.  
However, this idealised shape is modified due to the particle-finding algorithm that introduces inefficiencies for $\Delta T<10$\,ns, while the requirement for a local maximum within $\pm 3$\,ns implies no two particles have $\Delta T$ below this value.
To search for modularity in this plot, the data were binned in steps of 10\,ps and
the autocorrelation function, 
$
R(\tau)=\frac{1}{N-\tau}\sum_{i=1}^{N-\tau} y_iy_{i+\tau}
$
was calculated, where $y_i$ is the number of entries in bin $i$.
The coarse-grained structure of $\Delta T$ results in a roughly exponential shape to $R(\tau)$, superimposed on top of which is a fine-grained sinusoidal pattern.
Fig.~\ref{fig:pico} (right) shows $R$ as a function of $\tau$, after the coarse-grained structure has been subtracted.
The periodic structure is fitted with a modulated sine function for which the period is determined to be $346\pm 3$\,ps, consistent with the operating frequency of the linac of 2856\,MHz.
Thus, in addition to observing the large-scale $\sim3\,\mu$s-wide structure of the linac pulses that approximate to, but are not precisely square waves, the detector is capable of resolving the individual 350\,ps-wide sub-pulses in the beam.

\section{Conclusions}
\label{sec:conclude}
The results of the beam test at Saint Luke's hospital, Dublin provide a proof of concept for the use of new generation fast detectors for medical applications.
Our results for charged particles show a similar response to standard ion chamber measurements but with an active area a factor twenty smaller and a time resolution a factor one hundred million better.
Using this new detector we are able to observe the time profile of the charged particles in the linac beam and detect the pulse modulation due to the RF.
We believe this is the first time either structure has been measured with a particle detector.

Studies of single particle interactions in high rate medical accelerators cannot be performed with standard diagnostic tools. 
The capabilities of a LGAD read out with the front-end electronics developed at the University of Kansas, show unprecedented performance in beam monitoring for radiotherapy machines, and the same strategy of data taking and analysis can be applied for hadron therapy machines.
The much shorter timescale of operation of LGADs suggest a promising alternative for precise dose measurements and has particular applicability when a rapid response is required, e.g.\ for fast profile measurements or scanning of moving beam profiles.

LGAD technology is still in its infancy and many improvements are to be expected.
More sophisticated electronics should make it possible to reduce the pulse-length by a factor of ten and this will allow an order of magnitude improvement in the single-particle rates that it can measure.
LGAD sensors and the associated electronics currently only operate with a few pads or strips, with current implementations up to 16 channels.
However, the next generation of particle physics experiments have plans for arrays of these sensors read-out using a front-end ASIC~\cite{CARTIGLIA2020164383}, and there are already prototypes of sensors with 512 pads of 1.3x1.3\,mm$^2$~\cite{CMS:2667167}.

As the field of medical physics is currently moving towards increasing the amount of radiation delivered per unit of time with the development of initiatives like Flash-Radiotherapy linear accelerators, the introduction of fast detectors becomes important: 
LGADs represent a promising option for beam monitoring and dosimetry as they combine excellent spatial and temporal precision.

\bibliography{Bibliography}

\end{document}